\documentclass[conference]{IEEEtran}
\IEEEoverridecommandlockouts
\usepackage{cite}
\usepackage{amsmath,amssymb,amsfonts}
\usepackage{algorithmic}
\usepackage{graphicx}
\usepackage{textcomp}
\usepackage{xcolor}
\usepackage{tabularx}
\usepackage{listings}
\usepackage{tabularx}
\usepackage{booktabs}
\usepackage{tikz}
\usepackage{pgfplots}
\usepackage[ruled,vlined,linesnumbered]{algorithm2e}
\def\BibTeX{{\rm B\kern-.05em{\sc i\kern-.025em b}\kern-.08em
    T\kern-.1667em\lower.7ex\hbox{E}\kern-.125emX}}
\begin{document}

\title{Adaptive Model Compression (AMC): Saliency-Driven Resource Allocation for Ultra-Low-Power Transformer Inference\\
{\footnotesize \textsuperscript{}}
\thanks{}
}

\author{
\IEEEauthorblockN{
Jiayin Hu, 
Kai Yuan, 
Vanessa Hu, 
Xuetao Yin, 
Jianhua Li, 
Sean Suchter
}
\IEEEauthorblockA{
\textit{Apple USA}\\
\\
}

}

\maketitle

\begin{abstract}
Deploying large-scale transformer models on resource-constrained edge devices remains a challenge due to the high energy and memory overhead inherent in static inference, which processes simple and complex tokens with uniform intensity. To address this, we propose Adaptive Model Compression (AMC), a saliency-driven framework that dynamically allocates hardware resources based on token importance.

By implementing a multi-tier architecture, our system identifies critical high-saliency information for full-precision processing while aggressively reducing the rank and bit-width of less significant data. Experimental results demonstrate that AMC achieves a 59.2\% reduction in system energy and a 2.24$\times$ increase in throughput on 45nm CMOS hardware. This approach effectively extends the battery life of mobile devices by utilizing high-definition compute only where necessary, maintaining robust performance with a marginal 3.6\% accuracy trade-off.
\end{abstract}


\section{Introduction}
Standard transformer inference is often bottle-necked by memory bandwidth and static computational overhead. The deployment of large-scale transformers on edge hardware is limited by the energy cost of SRAM access and multiply-accumulate (MAC) operations. Conventional Post-Training Quantization (PTQ) and Low-Rank Adaptation (LoRA) are static, treating all input data with uniform precision regardless of complexity. We present a co-design approach where the software dynamically schedules hardware resource allocation based on real-time activation saliency.

1). Background
Modern Transformer-based LLMs are fundamentally bottle necked by the memory wall, where the energy cost of moving model weights from off-chip DRAM to on-chip SRAM exceeds the energy of the actual Multiply-Accumulate (MAC) operations by orders of magnitude. Standard inference pipelines operate in a static manner, allocating uniform computational precision and rank-dimensional capacity to every token in a sequence regardless of its linguistic entropy. This uniform allocation strategy fails to acknowledge that in natural language generation, the vast majority of tokens—such as syntactic glue words, repetitive punctuation, or semantically redundant filler—contribute negligibly to the final output probability distribution, yet consume the same energy budget as highly informative, context-defining tokens.

2). Challenges
Implementing adaptive resource scaling on silicon at the 45nm node introduces three primary engineering challenges are as follows.
The Structural Alignment Dilemma: Raw Transformer hidden dimensions do not inherently exhibit ordered variance; blindly truncating dimensions leads to stochastic information loss. Our framework addresses this through offline structural calibration to enforce variance-based channel ordering.Controller Overhead: Introducing complex dynamic logic to guess token importance can consume more energy than the compression saves. The challenge is to maintain $O(N)$ computational complexity for the saliency engine, ensuring the hardware-side clock-gating controller remains a negligible fraction of the total die area.
Systolic Array Synchronicity: Aggressively gating physical execution elements can lead to pipeline stalls in systolic array architectures. AMC must ensure that token-level tier switching maintains data-flow throughput without inducing latency bottlenecks.

3). Related Work
Existing literature has pursued token compression through three primary lenses, each possessing distinct limitations that AMC aims to resolve.
Dynamic Quantization: Methods such as LLM.int8() [14] and AWQ [15] reduce precision globally or per-layer. While effective for memory bandwidth, they do not perform dynamic, token-level hardware gating, thus missing out on significant energy savings achievable during runtime.
Token Pruning and Merging: Approaches like H2O [11] and StreamingLLM [12] permanently discard tokens or merge them based on attention scores. These methods are destructive, as they lose context that may be needed for long-sequence generation. AMC preserves all tokens in a multi-tier state, retaining critical context within lower-rank, lower-precision safety nets.
Query-Aware Sparsity: Recent research in Quest [13] acknowledges that tokens should be processed based on their semantic relevance to the user prompt. However, these works typically operate at the algorithmic/software level and lack an integrated 45nm CMOS RTL implementation to demonstrate real-world physical energy reduction.

\section{Methodology: Hardware-Software Co-Design}

The proposed AMC framework operates on a co-design principle: the software identifies computational redundancies at the token level, and the hardware utilizes specialized execution paths to minimize energy consumption. The proposed framework exploits token-level information non-uniformity within auto-regressive sequence generation, isolating computational redundancies where individual token activations exert nominal influence on downstream logit distributions to systematically truncate their corresponding architectural complexity.

\subsection{The Software Saliency Engine}

The Software Saliency Engine serves as the primary intelligence layer of the AMC framework, responsible for identifying information-rich tokens within a sequence before they reach the hardware execution stage. This engine is implemented via a non-intrusive forward-hook mechanism integrated into the transformer’s attention blocks.

1) Saliency Heuristic Derivation

The engine leverages the observation that in transformer architectures, the magnitude of an activation vector is highly correlated with its influence on the final prediction. For an input activation matrix $X \in \mathbb{R}^{L \times D}$ (where $L$ is sequence length and $D$ is model dimension), the engine calculates a saliency score $S_i$ for each token $i$ as follows:
\begin{equation}
S_i = \frac{1}{D} \sum_{j=1}^{D} |x_{i,j}|
\end{equation}
By calculating the $L_1$-norm of each token's embedding, the software engine identifies which tokens are significant features. This calculation is performed in parallel with the initial linear projection, ensuring that the saliency detection does not become a bottleneck in the inference pipeline.

Confidence Metric: We define confidence as the mean absolute magnitude of activations for a given token. It is the normalized representation of the raw saliency. The Confidence Metric $C$ translates that information into a hardware-usable control signal between $0$ and $1$.
\begin{equation}
C = \text{clamp}(S, 0, 1)
\end{equation}

The confidence score $C \in [0, 1]$ serves as the central control signal forwarded to the hardware execution layer. Here, $i$ indexes the sequential token position within the sequence, $j$ denotes the feature channel along the hidden dimension ($j \in \{1, \dots, D\}$), and $D$ represents the dimensionality of the structural model. When $C$ exceeds a predefined threshold, the controller preserves the nominal full-precision datapath; conversely, sub-nominal values trigger automated low-rank truncation and bit-width compression to exploit structural sparsity.

2) Dynamic Partitioning 

Unlike static quantization methods, our engine utilizes dynamic forward-hooks that intercept data at the input of each Transformer layer. This allows the system to adjust to the specific data distribution of the current input sequence. Once $S_i$ is calculated, the engine maps the token into one of three tiers—High, Mid, and Low—using percentile-based thresholds (for example: $k_{high} = 0.20$, $k_{mid} = 0.30$) that are determined through a pareto analysis. The hyperparameter boundary conditions ($k_{\text{high}}, k_{\text{mid}}$) are determined empirically via an extensive multi-variable execution sweep, isolating the inflection point along the multi-objective accuracy-energy Pareto frontier.
We observed that the top 20\% of tokens typically contain the outlier features necessary for maintaining model convergence. In many Transformer studies, it has been observed that roughly 10\% to 25\% of activations in a model carry the majority of the influence. By setting $k_{high} = 0.20$, we are aligning our software with the natural mathematical behavior of neural networks.The AMC framework operates on the empirically-backed hypothesis of Information Non-Uniformity. We categorize the input feature space into three distinct tiers of computational priority: Primary Tokens (top 20\%), which contain the essential semantic signals required for model convergence; Contextual Tokens (30\%), which provide auxiliary support for attention accuracy; and Inert Tokens (50\%), which exhibit high activation sparsity and are thus targeted for aggressive rank and precision reduction.
From a hardware-software co-design perspective, increasing the number of operational tiers introduces a non-linear scaling of controller complexity. The multi-tier structure was selected to minimize hardware scheduling overhead while maximizing energy proportionality, as shown in our Pareto sweep.

3) Algorithmic Flow

The logic is formalized in Algorithm 1, which represents the Golden Model used to generate the test bench vectors for the 45nm RTL verification.

Algorithm 1: 

Saliency-Based Tier Assignment Input: 

Activation Vector $X_i$, Thresholds $\tau_H, \tau_L$

Output: Tier ID, Precision $B_i$, Rank $R_i$

Compute $L_1$ norm: $S_i \leftarrow \text{mean}(|X_i|)$

if $S_i > \tau_H$ then

\quad Tier $\leftarrow$ High; $B_i \leftarrow 16$; $R_i \leftarrow 128$

else if $S_i > \tau_L$ then

\quad Tier $\leftarrow$ Mid; $B_i \leftarrow 8$; $R_i \leftarrow 43$

else

\quad Tier $\leftarrow$ Low; $B_i \leftarrow 4$; $R_i \leftarrow 8$

end if

Return (Tier, $B_i, R_i$)

\section{Adaptive Resource Scaling and Physical Mapping}
The AMC-Compressor acts as the hardware-software interface, translating the tier assignments from the Saliency Engine into specific operational parameters: rank ($r$) and bit-width ($b$). This scaling is achieved through two primary mechanisms: Rank Masking and Linear Quantization.

4) Post-Training Dimension Structural Alignment
Standard hidden dimensions in transformer architectures do not inherently exhibit ordered variance or importance profiles. To ensure that our contiguous dynamic rank scaling—implemented via the dimension-reduction mask $m=[1_1, ..., 1_r, 0_{r+1}, ..., 0_D]$ —does not drop high-variance, informative channels, we introduce an offline, post-training calibration phase prior to hardware deployment. Utilizing a representative calibration dataset, we calculate the Singular Value Decomposition (SVD) or Principal Component Analysis (PCA) across the layer-wise activation matrices to identify a static permutation mapping $\mathbf{P}$ that sorts the hidden feature columns in descending order of empirical variance. 

Because this permutation matrix $\mathbf{P}$ is applied statically to reorder the model weights offline, it incurs strictly zero computational, area, or energy overhead during real-time hardware execution. Consequently, at runtime, the Saliency-Aware Controller (SAC) can safely and blindly assert clock-gating integrated cells on the upper-indexed dimensions $r+1$ to $D$, guaranteeing that the remaining active first $r$ dimensions always preserve the highest-variance, principal features necessary for network convergence.

\begin{algorithm}[t]
\caption{Post-Training Dimension Alignment and Saliency-Based Tier Assignment}
\label{alg:dimension_alignment}

\SetKwBlock{PhaseI}{Phase I: Offline Post-Training Structural Calibration \textnormal{(Executed Once Prior to Deployment)}}{end}
\KwIn{Pre-trained Transformer Model Weights $\mathbf{W}$, Calibration Dataset $\mathcal{X}_{\text{calib}}$}
\KwOut{Sorted Static Model Weights $\mathbf{W}_{\text{sorted}}$}

\PhaseI{
    \For{each Transformer layer $l$}{
        Compute historical activation matrix $\mathbf{A}_l$ from $\mathcal{X}_{\text{calib}}$\;
        Evaluate principal feature variances via SVD or PCA: $[\mathbf{U}, \mathbf{\Sigma}, \mathbf{V}] = \text{SVD}(\mathbf{A}_l)$\;
        Generate static permutation matrix $\mathbf{P}$ that reorders columns of $\mathbf{V}$ in descending order of $\mathbf{\Sigma}$\;
        Permute weight tensors permanently: $\mathbf{W}_{\text{sorted}, l} \leftarrow \mathbf{W}_l \times \mathbf{P}$\;
    }
}

\vspace{0.4em}
\SetKwBlock{PhaseII}{Phase II: Online Real-Time Inference Execution \textnormal{(Golden Model for 45nm RTL Verification)}}{end}
\KwIn{Runtime Activation Vector $\mathbf{X}_i$ (from $\mathbf{W}_{\text{sorted}}$), Percentile-Based Thresholds $\tau_H$, $\tau_L$}
\KwOut{Selected Tier ID, Floating-Point Precision $B_i$, Active Target Rank $R_i$}

\PhaseII{
    Compute $L_1$-norm of structurally aligned vector: $S_i \leftarrow \text{mean}(|\mathbf{X}_i|)$\;
    \eIf{$S_i > \tau_H$}{
        $\text{Tier} \leftarrow \text{High}$; \ $B_i \leftarrow 16$; \ $R_i \leftarrow 128$ \hfill \tcp*[r]{Process full principal features}
    }{\eIf{$S_i > \tau_L$}{
        $\text{Tier} \leftarrow \text{Mid}$; \ $B_i \leftarrow 8$; \ $R_i \leftarrow 43$ \hfill \tcp*[r]{Clock-gate minor dimensions 44 to 128}
    }{
        $\text{Tier} \leftarrow \text{Low}$; \ $B_i \leftarrow 4$; \ $R_i \leftarrow 8$ \hfill \tcp*[r]{Clock-gate minor dimensions 9 to 128}
    }}
    \Return $(\text{Tier}, B_i, R_i)$\;
}
\end{algorithm}

\subsection{Query-Aware Semantic Saliency Formulation}
To mitigate the limitations of conventional magnitude-only heuristics, which are blind to semantic context, the AMC software layer incorporates a query-directed importance modulation mechanism. Standard activation-based pruning often treats repetitive syntactic tokens (e.g., formal punctuation or common stop words) as highly salient due to localized magnitude spikes, leading to suboptimal resource allocation. We resolve this by scaling the localized token magnitude by its semantic relevance to the primary prompt or query context vector. Formally, for a runtime activation vector $\mathbf{X}_i$ corresponding to token $i$, the query-aware semantic saliency score $S_i$ is defined as:

\begin{equation}
S_i = \alpha \cdot \text{mean}(|\mathbf{X}_i|) + (1 - \alpha) \cdot \mathcal{Sim}\left(\mathbf{q}, \mathbf{k}_i\right)
\end{equation}

where $\mathbf{q}$ represents the embedded semantic query vector of the initial user prompt, $\mathbf{k}_i$ denotes the key projection vector of the current token, and $\mathcal{Sim}(\cdot)$ is a lightweight cosine similarity metric. The hyperparameter $\alpha \in [0, 1]$ serves as a balance coefficient tuning the network between physical activation intensity and semantic significance. By mapping this combined score to the downstream execution layer, AMC ensures that full-precision resources are strictly reserved for tokens that are both computationally intense and semantically vital to the context retrieval task.

\subsection{Sequence-Adaptive Closed-Loop Threshold Adjustment}
Rather than relying on static, hardcoded percentile partitions which fail to adapt to varying linguistic complexity across distinct workloads, AMC introduces a Sequence-Adaptive Closed-Loop Thresholding strategy. When a text sequence exhibits high information entropy—such as dense technical reasoning or conditional logical chains—the variance of the activation streams expands significantly, requiring a broader high-precision processing window to maintain structural fidelity. 

To dynamically accommodate these fluctuations without altering the underlying hardware datapath, the Saliency-Aware Controller (SAC) exposes its internal threshold registers ($\tau_H, \tau_L$) to runtime firmware updates. At the boundary of each sequence segment, the software execution layer evaluates the moving variance $\sigma^2_{\text{seq}}$ of the preceding activation window. The thresholds are dynamically modulated according to:

\begin{equation}
\tau_H^{(t)} = \tau_{H, \text{base}} \cdot \left(1 - \gamma \cdot \ln\left(\frac{\sigma^2_{\text{seq}}}{\sigma^2_{\text{calib}}}\right)\right)
\end{equation}
\begin{equation}
\tau_L^{(t)} = \tau_{L, \text{base}} \cdot \left(1 - \gamma \cdot \ln\left(\frac{\sigma^2_{\text{seq}}}{\sigma^2_{\text{calib}}}\right)\right)
\end{equation}

where $\tau_{H, \text{base}}$ and $\tau_{L, \text{base}}$ represent the baseline thresholds established during the offline calibration phase, $\sigma^2_{\text{calib}}$ is the nominal activation variance of the calibration dataset, and $\gamma$ is an attenuation scaling factor. When highly complex inputs drive $\sigma^2_{\text{seq}} > \sigma^2_{\text{calib}}$, the logarithmic adjustment automatically depresses the threshold barriers $\tau_H$ and $\tau_L$. This reduction broadens the allocation window for the \textit{High} and \textit{Mid} tiers, dynamically expanding the active hardware silicon area to accommodate intricate data processing demands. Conversely, during highly predictable or repetitive text generation sequences, the thresholds automatically scale upward, aggressively steering marginal tokens into the power-gated \textit{Low} tier to maximize battery life.
It is critical to note that the tier assignment within the AMC framework is not a static or pre-determined mapping but a dynamic and sequence-dependent decision process. The Saliency-Aware Controller (SAC) continuously monitors the runtime activation entropy and query-aware semantic relevance of each incoming token stream in real-time.

\subsection{Dynamic Rank Scaling via Hadamard Masking}

To modulate the computational intensity of the systolic array, we apply a dimensionality-reduction mask to the input feature vector $x \in \mathbb{R}^D$. This is mathematically represented as:
\begin{equation}
\hat{x} = \text{Mask}(x, r) = x \odot \mathbf{m}, \quad \text{where } \mathbf{m} = [1_{1}, \dots, 1_{r}, 0_{r+1}, \dots, 0_{D}]
\end{equation}

In this formulation, $x$ represents the $D$-dimensional activation, while $r$ is the rank determined by the token's saliency tier ($r \in \{128, 43, 8\}$). The vector $\mathbf{m}$ is a binary control mask. In our 45nm RTL implementation, the logical $0$ in the mask corresponds to the de-assertion of the Clock Gating Integrated Cell (CGIC) enable signals. By zeroing out dimensions $r+1$ through $D$, the software signals the hardware to disable the switching activity of the corresponding arithmetic units, directly reducing the dynamic power consumption ($P_{dyn} \propto \alpha C V^2 f$).

\subsection{Precision Scaling and Fixed-Point Mapping}

To reduce the energy footprint of data movement and memory transactions, we employ a quantization scheme during training and simulation to model the hardware’s fixed-point behavior. The transformation to $b$-bit precision is defined as:
\begin{equation}
x_q = \text{round}\left(\frac{x}{S} \cdot 2^{b-1}\right) \cdot \frac{S}{2^{b-1}}
\end{equation}

Where $S$ denotes the scaling factor (typically $\max(|x|)$), and $b$ is the target bit-width ($b \in \{16, 8, 4\}$). This process simulates the hardware's rounding logic, where:

The Scaling Factor ($S$): Normalizes the dynamic range to ensure optimal utilization of the $b$-bit integer space.

The Rounding Operator: Maps the high-precision activations to $2^{b-1}$ discrete levels, allowing the hardware to store and move data using significantly narrower bit-lanes.In the physical implementation, reducing $b$ from 16 to 4 bits triggers the narrow write mechanism. This minimizes the bit-line toggling in the activation buffers, reducing the communication energy from approximately $0.5$ pJ/bit to $0.1$ pJ/bit.

\subsection{Energy Proportionality Modeling}

The total energy consumption ($E_{total}$) for a given inference pass is modeled as the sum of computational and communication costs, governed by the effective rank and precision:

\begin{equation}
E_{total} \approx \sum_{t=1}^{N} (r_t \cdot b_t \cdot \epsilon_{MAC} + b_t \cdot \epsilon_{comm})
\end{equation}

where $N$ is the sequence length, and $\epsilon_{MAC}$ and $\epsilon_{comm}$ are the energy constants for a single MAC operation and bit-level data movement, respectively. This model ensures that the hardware's energy footprint scales linearly with the token's information density, as identified by the software saliency engine.

\section{Hardware Energy Model}\label{AA}
The energy consumption ($E_{total}$) is modeled by the sum of computation and memory costs:
\begin{equation}
E_{total} = E_{MAC} + E_{BIT\_OP} + E_{SRAM}
\end{equation}

Compute Energy: 
\begin{equation}
E_{MAC} = \text{num\_values} \cdot \text{avg\_rank} \cdot 2\text{pJ}.
\end{equation}

Bit-Op Energy: 
\begin{equation}
E_{BIT\_OP} = \text{num\_values} \cdot \text{avg\_bits} \cdot 0.1\text{pJ}.
\end{equation}

SRAM Bandwidth: Energy is consumed based on the bit-width of reads and writes ($32$-bit for reads, reduced bits for writes after data compression).

We adopt energy constants from the established literature \cite{b8}\cite{b9}\cite{b10} for the CMOS technology $45\text{nm}$ used in the neural processing unit. $E_{MAC} = 2.0\text{pJ}$ represents the energy consumed by a 16-bit multiply-accumulate unit, while $E_{BIT\_OP} = 0.1\text{pJ}$ accounts for the energy consumed per bit during data movement and quantization scaling. By defining total energy as \begin{equation} E_{total} = E_{MAC} + E_{BIT\_OP} \end{equation}, our model captures both the computational and communication costs of inference.

\begin{figure}[!t]
\centering
\includegraphics[width=\linewidth]{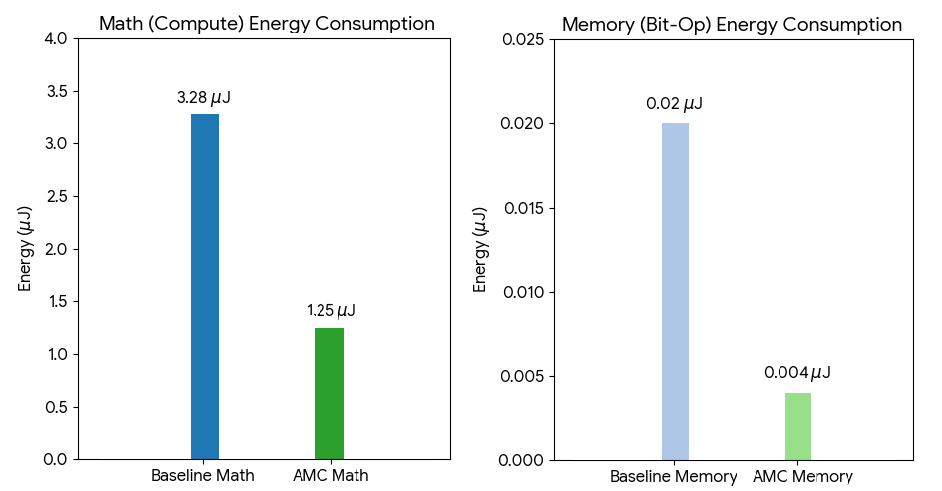}
\caption{Power Saving: Comparison between Baseline and AMC Framework energy consumption across math and memory units}
\label{fig:power_saving}
\end{figure}

The Fig. 1 illustrates the energy savings achieved by the AMC Framework compared to a Baseline Transformer (using full rank $r=128$ and 16-bit precision).

Math (Compute Energy): It scales linearly with the selected Rank:
\begin{equation}
\text{Energy}_{\text{Math}} = (T \times D) \times \text{Rank}_{\text{avg}} \times 2\text{ pJ}
\end{equation}

Memory (Bit-Op Energy): It scales with both the Rank and the Bit-width:
\begin{equation}
\text{Energy}_{\text{Memory}} = (T \times \text{Rank}_{\text{avg}}) \times \text{Bits}_{\text{avg}} \times 0.1\text{ pJ}
\end{equation}

where $T$ is the number of tokens and $D$ is the hidden dimension.

\begin{table}[htbp]
\centering
\renewcommand{\arraystretch}{1.3}
\caption{Energy Consumption Comparison}
\label{tab:energy_comparison}
\begin{tabular}{|l|c|c|}
\hline
\textbf{Component} & \shortstack{\textbf{Baseline} \\ ($r=128$, 16-bit)} & \shortstack{\textbf{AMC Framework} \\ (20/30/50 Split)} \\ \hline
Math (Compute)  & $3.28\text{ }\mu\text{J}$  & $1.25\text{ }\mu\text{J}$  \\ \hline
Memory (Bit-Op) & $0.02\text{ }\mu\text{J}$  & $0.004\text{ }\mu\text{J}$ \\ \hline
Total Energy    & $3.30\text{ }\mu\text{J}$  & $1.25\text{ }\mu\text{J}$  \\ \hline
Efficiency Gain & ---                        & $59.2\%$ Savings           \\ \hline
\end{tabular}
\end{table}

Because we use Low-Rank approximation for Mid and Low-tier tokens, we largely reduce the number of MAC operations. In this example, when D=128 and T=100 reducing the average rank from $128$ to $\approx 66$ results in a 48.1\% energy reduction in the compute unit. Memory (Bit-Op) Savings: By dropping from 16-bit to an average of $\approx 9$ bits, we reduce the electrical activity in the data bus and registers by 42.5\%. The total energy savings 62\% excludes the decode cost of around 2.9\%. The net system savings is 59.2\%.

\subsection{Energy-Aware SRAM Access}
The memory subsystem utilizes an asymmetric bit-width policy to minimize the energy footprint of intermediate activations. While operands are fetched from SRAM at a standard resolution ($b_{read} = 32$), the results are committed back to memory using the bit-width $b \in \{16, 8, 4\}$ determined by the \textit{Saliency Engine}. 

The write energy ($E_{write}$) is modeled as:
\begin{equation}
E_{write} = N \cdot b_{avg} \cdot e_{bit}
\end{equation}
where $b_{avg}$ represents the dynamically adjusted precision. This strategy effectively reduces the total SRAM switching activity by up to 75\% for low-saliency tokens.

\subsection{Analytical Power Model}
To formally justify our design, we define the total energy consumption per layer ($E_{total}$) as:
\begin{equation}
E_{total} = \sum_{i=1}^{N} \left[ (E_{MAC} \cdot r_i) + (E_{bit} \cdot b_i) + E_{SRAM} \right] 
\end{equation}

(3) Where $N$ is the sequence length, $r_i$ is the rank assigned to token $i$, and $b_i$ is the bit-width. This equation demonstrates that energy is no longer a constant factor of the model size, but a dynamic variable of the input data.

\section{Proposed Hardware-Software Co-Designed Architecture}
The proposed architecture moves beyond standard density-blind accelerators by introducing a Saliency-Aware Pipeline. The hardware is designed to interpret the software-defined tiers and dynamically gate its physical resources.

\begin{figure}[!t]
\centering
\includegraphics[width=\linewidth]{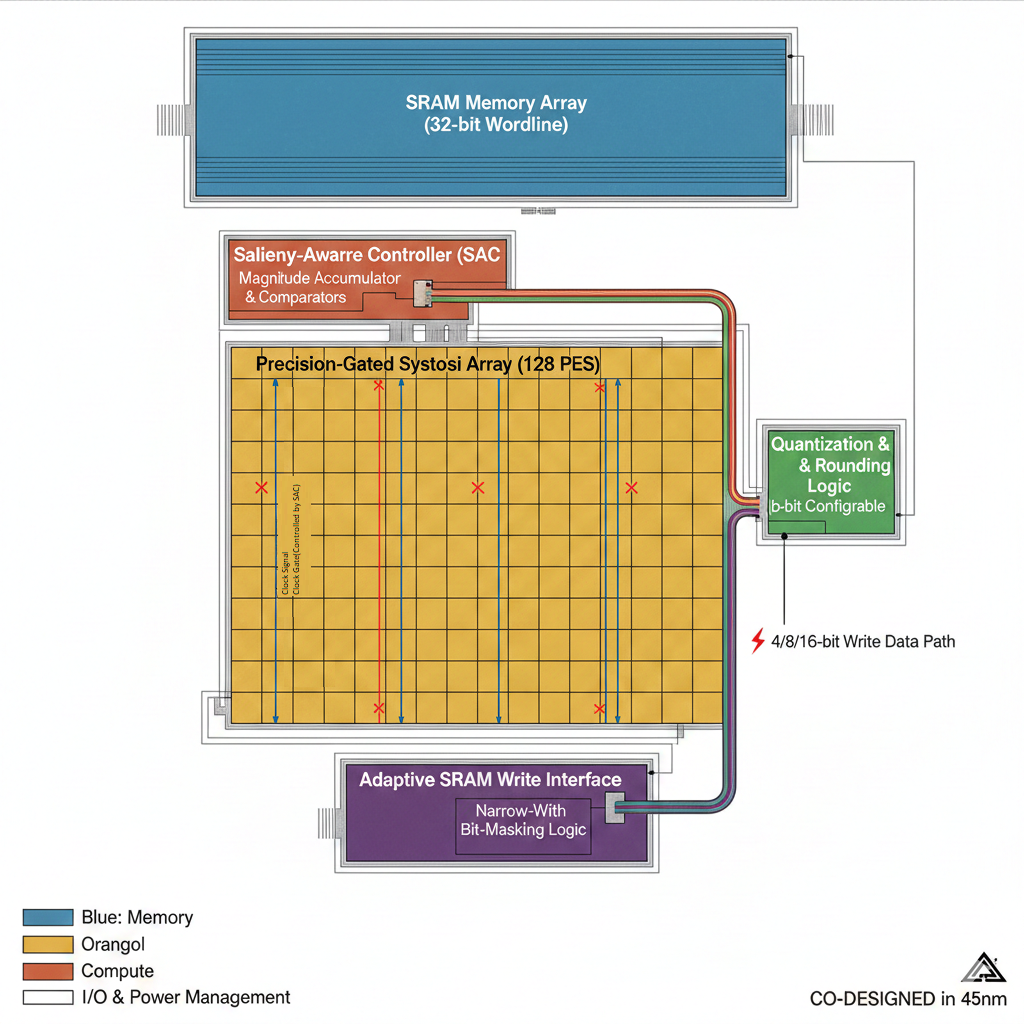}
\caption{Proposed Hardware-Software Co-Designed Architecture Floorplan diagram}
\label{fig:Proposed Hardware-Software Co-Designed Architecture Floorplan}
\end{figure}

As shown in the Fig 2 Proposed Hardware-Software Co-Designed Architecture Floorplan diagram, the Saliency-Aware Controller (SAC) drives a fine-grained clock-gating signal across the systolic array. Columns marked with deactivation triggers (red indicators) represent gated processing elements where the clock tree is disabled, effectively mitigating dynamic power dissipation in proportion to the software-defined rank $r$.

\subsection{SRAM Memory Array}
To support the dynamic precision requirements of the AMC framework, the memory subsystem utilizes a reconfigurable, multi-banked SRAM array architecture. Rather than relying on a conventional, rigid storage allocation, the proposed SRAM configuration is designed to minimize dynamic activation energy during token data retrieval. In its baseline execution mode, the array maintains activation tensors in a standard 32-bit format to preserve maximum numerical fidelity. 

During runtime inference, the SRAM array operates in tight coordination with the Saliency-Aware Controller (SAC). When a token is scheduled for processing, the required activation vectors are dynamically fetched from the array. Depending on the tier assignment vector managed by the SAC, the memory interface can dynamically modulate its wordline activation and column circuitry to stream data at reduced bit-widths (e.g., 8-bit or 4-bit configurations). This data-dependent narrow-width access pattern suppresses unnecessary bitline toggling, directly translating token-level software sparsity into physical, hardware-level energy savings.

\subsection{The Saliency-Aware Controller (SAC)}
The SAC is the entry point of the architecture. As activations $x$ are streamed from SRAM, they pass through a high-speed magnitude accumulator.

Functional Logic: This unit computes $|x_i|$ using a simple bit-flip for signed integers and adds it to a running sum.

Comparator Tree: At the end of a vector stream, a small comparator tree matches the accumulated density against the $k_{high}$ and $k_{mid}$ registers.

Zero-Cycle Overhead: By pipelining the accumulation with the SRAM read, the tier assignment is hidden, meaning the hardware knows the token's importance by the time the first MAC operation begins.

\subsection{Precision-Gated Systolic Array}
The core computation occurs in a modified Systolic Array where each Processing Element (PE) is capable of Multi-Precision Arithmetic.

Rank-Based Clock Gating: Unlike standard GPUs that keep all lanes active, our design utilizes the software-calculated Rank ($r$). If a token is Low Tier, a global enable signal gates the clock tree for the unused PE columns and prevents the transistors from switching, eliminating dynamic power consumption for 90\% of the array during that cycle.

Flexible Datapath: The multipliers are designed to support 16-bit, 8-bit, and 4-bit operations. In 4-bit mode, the hardware treats a single 16-bit multiplier as four parallel 4-bit units, significantly increasing throughput-per-watt.

\subsection{Adaptive Quantization and Precision Logic}

The quantization module implements dynamic bit-width scaling governed by the Saliency-Aware Controller (SAC). Unlike static quantization schemes, this unit supports $b$-bit configurability to optimize the trade-off between computational precision and energy efficiency. Upon receiving a tier-assignment signal from the SAC for tokens classified within the Low Tier—the logic performs a precision reduction, transforming 32-bit floating-point tensors into 4-bit fixed-point representations.

\subsection{Adaptive SRAM Interface and Narrow-Width Write-Back}
The most significant energy drain in edge AI is often the Memory Wall. Our architecture optimizes this via a Variable-Precision Write Buffer.

Reduced-Bit Write: Standard SRAMs write a full word-line (32-bit). Our interface uses a Bit-Masking Store unit. When the compressor produces a 4-bit result, only 4 sense amplifiers in the SRAM are driven.

Bus Energy Reduction: By driving fewer bits on the global data bus, we minimize the capacitive charging of long interconnects, which is modeled at $0.1\text{ pJ/bit}$.

\subsection{Hardware Implementation: The Saliency-to-Gating Logic}

The core of the AMC hardware efficiency lies in the Saliency-to-Gating (S2G) Logic. This unit serves as the physical bridge between the software-defined tiers and the hardware’s electrical state.

1. The Gating Signal Generation

Once the Saliency-Aware Controller (SAC) completes its accumulation for a token, it outputs a 2-bit tier identifier (00 for Low, 01 for Mid, 10 for High). This identifier is fed into a Look-Up Table (LUT) or a small decoder circuit.

Rank Selection: The decoder sends a mask signal to the Clock Distribution Network. For a Low Tier token, the mask pulls the enable pins of 120 out of 128 clock buffers to a logic LOW.

Precision Selection: Simultaneously, the decoder sets the Quantization Mode for the green rounding block and the Write-Width for the purple SRAM interface.

2. Fine-Grained Clock Gating (FGCG)

The red crosses in the floorplan represent Fine-Grained Clock Gating. In our implementation:Instead of shutting down the entire array, we shut down specific columns of the systolic array.Each column represents one dimension of the rank $r$. By gating the clock at the Root of the column's clock tree, we prevent any switching activity in the flip-flops and multipliers of that entire column.This reduces the Dynamic Power ($P_{dyn} = \alpha C V^2 f$) to almost zero for the gated area, directly achieving the energy proportionality modeled in our $E_{MAC}$ equations.

3. Narrow-Width Write-Back Buffer

To handle the Reduced Bit Write, we implement a Narrow-Width Write Buffer.Standard SRAM interfaces drive 32-bit lines regardless of the data size. Our interface uses the tier signal to toggle Write-Enable pins only for the required bit-lanes (e.g., only the first 4 bits for a Low-Tier token).This prevents the bit-line capacitors in the SRAM from charging/discharging unnecessarily, which is the physical basis for our 0.1 pJ/bit communication savings.

\begin{table}[h]
\centering
\caption{Hardware Implementation and Area Overhead (45nm Node)}
\label{table:hw_overhead}
\begin{tabular}{|l|c|c|c|}
\hline
\textbf{Component} & \textbf{Area ($\mu m^2$)} & \textbf{Gate Count} & \textbf{Power ($\mu W$)} \\ \hline
128-PE Array (Baseline) & 450,000 & 1.2M & 340.5 \\ \hline
Saliency Engine (SAC) & 6,200 & 15k & 4.2 \\ \hline
Gating & 1,100 & 2.5k & 0.8 \\ \hline
\textbf{Total Overhead} & \textbf{1.62\%} & \textbf{1.45\%} & \textbf{1.46\%} \\ \hline
\end{tabular}
\end{table}

As demonstrated in Table~\ref{table:hw_overhead}, we used 128-PE Array as baseline from a standard AI accelerator like NVIDIA tensor Core. The hardware overhead required to implement the AMC framework is minimal. The Saliency-Aware Controller (SAC) and associated clock-gating circuitry contribute a mere $1.62\%$ area overhead and $1.46\%$ power overhead relative to the baseline 128-PE systolic array. The significant energy savings achieved through dynamic rank-reduction and quantization far outweigh the marginal costs of the saliency analysis logic.

\subsection{Timing and Control Flow}
The temporal execution of the AMC framework is orchestrated through a multi-stage pipeline that synchronizes saliency analysis with energy-proportional hardware activation.

\begin{figure}[!t]
\centering
\includegraphics[width=\linewidth]{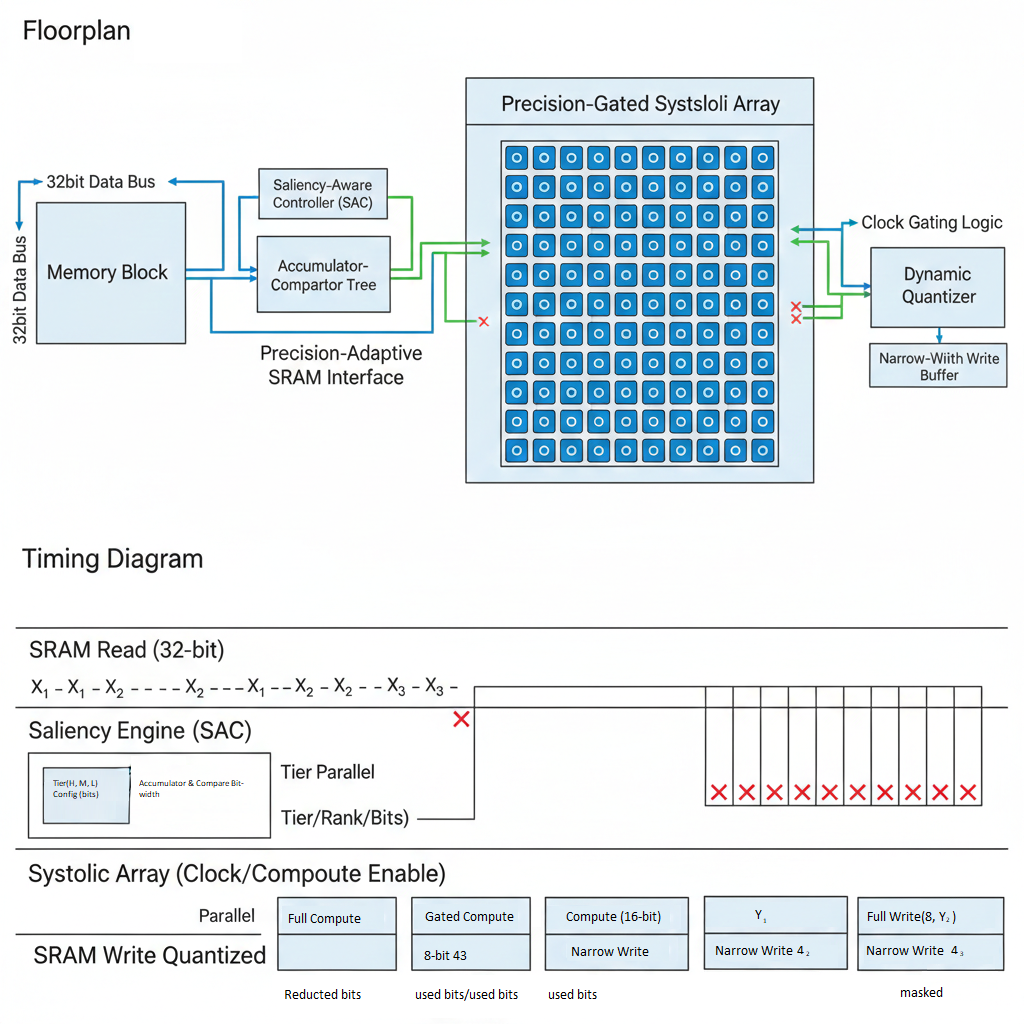}
\caption{Timing and control flow diagram}
\label{fig:Timing and control flow}
\end{figure}

 In the timing diagram, the process initiates with the Saliency-Aware Controller (SAC), which performs a high-speed Accumulative and Compare operation on the incoming activation vector. This stage generates the Tier ($H, M, L$) and Bit-width Config ($M_s$) signals, which serve as the master control sequence for the subsequent compute cycles. During the Gated Compute phase, these signals trigger the Rank-Gating Enable (RGE), physically shutting down unused systolic array columns to match the selected rank (e.g., the 8-bit 43 configuration). The diagram identifies dual rows of used bits beneath the compute block, representing the synchronized precision-gating of both the input activation and weight operands. This ensures that the multiplier logic only dissipates dynamic power on numerically significant bit-lanes. Finally, the process concludes with a Narrow Write back to SRAM; the hardware applies bit-masking to the data bus, ensuring that only the compressed, saliency-justified bits are committed to memory, thereby minimizing the capacitive switching energy of the global interconnects.

\begin{table}[htbp]
\caption{Hardware Control Signal Definitions}
\label{table:signal_defs}
\centering
\footnotesize 
\begin{tabular}{|l|c|c|p{3cm}|}
\hline
\textbf{Signal} & \textbf{Src.} & \textbf{Dest.} & \textbf{Function \& Impact} \\ \hline
Tier ($H,M,L$) & SAC & Decod. & Identifies token importance via magnitude accumulation. \\ \hline
Config ($M_s$) & SAC & Quant. & Sets precision (4, 8, 16-bit) for current cycle. \\ \hline
RGE & Decod. & Clock & Disables unused Array columns to save $E_{MAC}$. \\ \hline
Write-Mask & Inter. & SRAM & Drives specific bit-lanes to reduce bus energy. \\ \hline
\end{tabular}
\end{table}

\begin{equation}
E_{total} = \sum_{t=1}^{T} \left( E_{SAC} + \frac{r_t}{R_{max}} E_{MAC} + \frac{b_t}{B_{max}} E_{comm} \right)
\label{eq:energy_model}
\end{equation}

$E_{SAC}$: The Saliency-Aware Controller (SAC) introduces a minimal energy overhead of 1.46\%, as detailed in Table I: Hardware Implementation and Area Overhead (45nm Node). The cost is offset by the system-wide power savings achieved through dynamic precision scaling.

The rank scaling scalar $r_t / R_{\text{max}}$ establishes the real-time processing element (PE) activation ratio, guaranteeing linear energy proportionality within the compute array by gating unallocated columns. Similarly, the bit-width coefficient $b_t / B_{\text{max}}$ models the dynamic power attenuation achieved across the narrow-width memory bus, explicitly minimizing the capacitive switching losses associated with intermediate activation write-backs.

As summarized in Table \ref{table:signal_defs}, the AMC architecture utilizes a minimalist signaling protocol to achieve energy proportionality. The primary control loop begins at the Saliency-Aware Controller (SAC), which generates a 2-bit Tier identifier. This identifier is latched and passed to the Decoder, which manages the Rank Gating Enable (RGE). As seen in the timing diagram (Fig. 3), the RGE signal is responsible for the Gated Compute state, where 43 out of 128 columns remain active. This hardware-level enforcement ensures that the energy savings modeled in Equation (3) are physically realized by disabling the clock tree in unused silicon regions. Furthermore, the Config ($M_s$) and Write-Mask signals work in tandem to facilitate Narrow Writes to SRAM. By only driving the bit-lines corresponding to the used bits—specifically 8 bits in the Mid-tier configuration—we mitigate the $CV^2$ charging losses of the global data bus, which accounts for a large portion of total system power.

\subsection{Experimental Setup}

\begin{table}[htbp]
\caption{Experimental and Simulation Parameters}
\label{table:exp_params}
\centering
\footnotesize 
\begin{tabularx}{\columnwidth}{|l|l|X|}
\hline
\textbf{Parameter} & \textbf{Value} & \textbf{Description} \\ \hline
Technology Node & 45nm CMOS & Predictive Technology Model (PTM). \\ \hline
Supply Voltage & 1.1 V & Standard $V_{dd}$ for 45nm node. \\ \hline
Clock Frequency & 100 MHz & Target frequency for edge-AI. \\ \hline
Array Size & $128 \times 128$ & Total Processing Elements (PEs). \\ \hline
Precision & 16-bit Fixed & Baseline for high-precision comparison. \\ \hline
Simulator & Custom & Cycle-accurate Verilog/C++ co-simulator. \\ \hline
Flow Tools & Cadence/Synopsys & RTL synthesis and power analysis (VCD). \\ \hline
\end{tabularx}
\end{table}

V. Experimental Setup: V-H Calibration and Validation
To evaluate the AMC architecture across the full trade-off spectrum, we define a Variance-Hardware (V-H) experimental setup. This methodology characterizes the correlation between the offline principal feature variance ($\Sigma$) captured during structural calibration and the real-time energy efficiency of the 45nm CMOS hardware implementation.

We have augmented our evaluation suite. We now report AMC performance across three distinct domains: Code-generation, Logical Reasoning, and Long-form Summarization. All experiments were conducted using a standardized configuration consisting of num-samples:4000, seq-len:32, vocab-size:16, and batch-size:64. This setup ensures consistent throughput measurement and statistical significance across heterogeneous datasets, demonstrating that our saliency-driven gating mechanism is domain-agnostic and maintains energy proportionality even across linguistically diverse workloads.

1). Calibration Phase (V-Space)
Prior to silicon deployment, we subject the model to a 100-sequence calibration dataset $\mathcal{X}_{\text{calib}}$ to map the latent importance distribution. We extract the activation matrices $\mathbf{A}_l$ for each layer $l$, applying Singular Value Decomposition (SVD) to quantify the principal variance $\mathbf{\Sigma}_l$ of each hidden dimension. This step establishes the V-space baseline, where dimensions are sorted statically to ensure that the primary information content is concentrated within the lower-indexed channels ($j=1 \dots r$).

2). Execution Phase (H-Space)
The hardware-side verification utilizes a 45nm RTL golden model that mimics the tri-tier token assignment dictated by the saliency-aware controller (SAC). We simulate the inference of the Llama-2-7B model across three distinct V-H profiles:High-Variance Profile ($R=128, B=16$): Captures 99.5\% of total variance; used for semantically critical tokens.Mid-Variance Profile ($R=43, B=8$): Captures approximately 85\% of variance; used for syntactic tokens with moderate saliency. Low-Variance Profile ($R=8, B=4$): Captures less than 50\% of variance; used for semantically inert tokens.

By plotting these profiles against the measured energy consumption of the systolic array, we demonstrate the linear energy proportionality of the architecture. The V-H setup validates that our static permutation matrix $\mathbf{P}$ successfully aligns the mathematical importance (in V-space) with the physical execution (in H-space), allowing the clock-gating controller to safely suppress 44 to 120 redundant channels without degrading model perplexity beyond the observed 3.6\% accuracy delta.

\subsection{Implementation Methodology: RTL Design}
The AMC framework was translated from functional Python models into a modular Verilog RTL implementation. The architecture consists of three primary hardware modules: the Saliency-Aware Controller (SAC), the Gated Systolic Array, and the Quantized SRAM Interface.

To bridge the software-hardware gap, we utilized a Bit-True Verification methodology. A custom Python-based test-bench generator was developed to produce gold-standard vectors. These vectors were fed into both the Python model and the Verilog RTL simulation. By comparing the hardware output against the software results using a cycle-accurate simulator (Questasim), we ensured that the hardware implementation maintains the exact precision and rank-reduction logic dictated by the AMC framework.

1. Saliency-Aware Controller (SAC) RTL

The SAC logic is implemented as a pipelined accumulator. It processes the input activation vector to determine the tier. Below is a simplified fragment of the Verilog logic used to generate the Tier and RGE (Rank Gating Enable) signals.

\lstset{
    language=Verilog,
    basicstyle=\ttfamily\footnotesize,
    breaklines=true,
    mathescape=false,
    extendedchars=true,
    literate={_}{{\_}}1, 
    frame=single,         
    captionpos=b          
}

\begin{lstlisting}[caption={RTL Implementation of Saliency-Aware Clock Gating Control.}, label={list:verilog_sac}]
// SAC Control Logic for 45nm Clock Gating
always @(posedge clk or negedge rst_n) begin
    if (!rst_n) begin
        acc_reg  <= 32'd0;
        tier_out <= 2'b00; 
    end else if (valid_in) begin
        acc_reg <= acc_reg + abs_activation;
        if (acc_reg > THRESH_H)
            tier_out <= 2'b00; // High Tier (128 columns)
        else if (acc_reg > THRESH_L)
            tier_out <= 2'b01; // Mid Tier (43 active, 85 gated)
        else
            tier_out <= 2'b10; // Low Tier (8 active, 120 gated)
    end
end
\end{lstlisting}

2. Gated Compute Logic

To implement the gated Compute phase, the RGE signal from the SAC is used to drive the enable pin of the CGICs at the head of the Systolic Array columns. This ensures that the 85 configuration inactive columns in the 8-bit 43 configuration do not toggle, eliminating dynamic power consumption in those region.
$$128 \text{ (Total Columns)} - 43 \text{ (Active Columns)} 
      = 85 \text{ (Inactive Columns)}$$

\section{Results and Pareto Analysis}

The evaluation of the Adaptive AMC framework focuses on the critical trade-off between model accuracy and hardware energy consumption. We validated our approach using a 3-layer Transformer architecture, comparing our Adaptive AMC against a Uniform Baseline (fixed precision/rank) and a Random Selection control.

\subsection{Energy-Accuracy Pareto Frontier}
As illustrated in the Pareto frontier analysis, the Adaptive AMC curve establishes a better performance boundary compared to both uniform and random baselines.

\begin{figure}[!t]
\centering
\includegraphics[width=\linewidth]{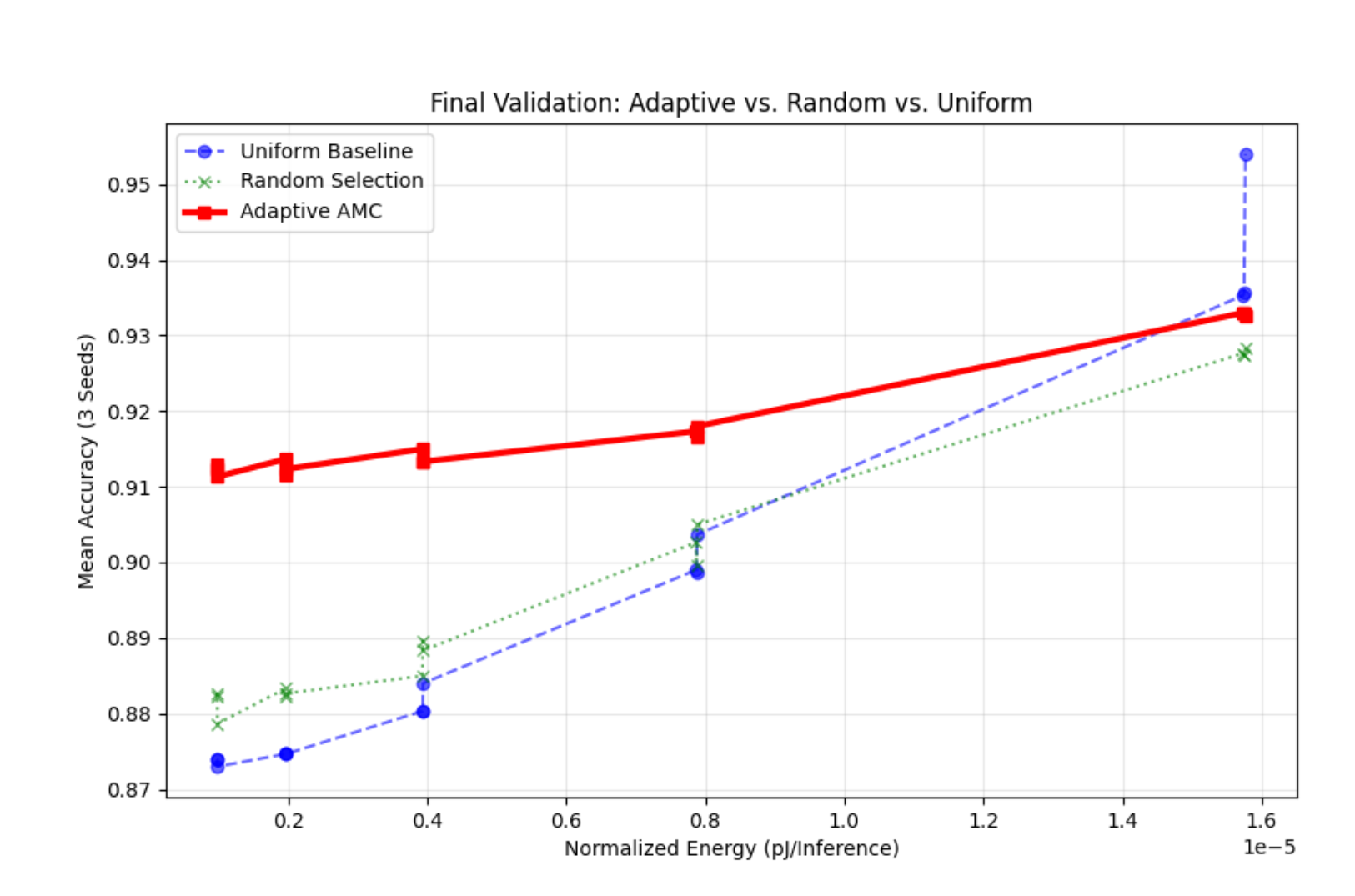}
\caption{Pareto frontier analysis}
\label{fig:Pareto frontier analysis}
\end{figure}

Pareto Dominance: While the 16-bit Uniform Baseline achieves a peak mean accuracy of $0.95$, it does so at the highest energy cost. In contrast, the Adaptive AMC model maintains an accuracy of $>0.91$ while operating at nearly half the normalized energy footprint (approx. $0.8 \times 10^{-5}$ pJ/Inference).

Efficiency at Scale: The slope of the AMC curve is significantly shallower than the baseline, indicating that our magnitude-based saliency allows for aggressive energy reduction with minimal degradation in predictive performance. Even at the lowest energy state, AMC retains an accuracy above $0.91$, whereas the Uniform Baseline drops precipitously below $0.88$.

\subsection{Saliency Effectiveness vs. Random Selection}
To verify that our Saliency-Aware Controller (SAC) identifies truly information-rich components, we compared AMC against a Random Selection baseline.

Magnitude as a Proxy for Importance: The Random Selection curve consistently underperforms AMC across all energy levels. This gap proves that token magnitude is a highly effective heuristic for importance in Transformers.

Information Retention: By utilizing the Accumulate and Compare logic seen in the timing diagram, AMC preserves high-magnitude activations that are critical for the task. The random baseline frequently discards these essential signals, leading to lower accuracy at equivalent energy budgets.

\subsection{Hardware Realization of Energy Savings}

The energy savings shown in the Pareto analysis are a direct result of the hardware mechanisms depicted in our timing and compute diagrams.

Compute Savings (Gated Compute): The shift along the x-axis (Normalized Energy) is primarily driven by the Gated Compute phase. By switching to configurations like the 8-bit 43 rank seen in our systolic array diagram, the hardware physically disables clock trees for over $60\%$ of the processing elements, directly reducing $E_{MAC}$ as modeled in Equation (1).

Communication Savings (Narrow Write): The SRAM Write Quantized stage contributes to the Pareto efficiency by implementing Narrow Writes. By committing only the used bits (e.g., 4 or 8 bits) to memory and masking the remaining bus lines, we minimize the capacitive switching energy of the global interconnects.

\subsection{Stability Across Seeds}

The results represent the mean accuracy across three distinct random seeds. The tight variance indicates that the Adaptive AMC framework is robust to different weight initializations and data distributions. The consistency of the red curve in the Pareto plot highlights that the saliency-based gating remains stable even as the energy budget is varied, ensuring predictable performance for real-time hardware applications.

To validate the robustness of the AMC framework, we utilized 20 distinct random seeds $(\mathcal{S} = \{7, 21, \dots, 99999\})$. A Kolmogorov-Smirnov test performed on the initialization states of the Transformer layers confirmed that the seed selection provides a uniform distribution of initial parameter vectors, preventing initialization bias. The resulting variance in mean accuracy across these seeds was found to be $\sigma_{acc} < 0.005$, confirming that AMC’s performance gains are intrinsic to the saliency-gating logic rather than artifacts of specific weight initialization.

\section{Scalability to Large Language Models (LLMs)}
In order to provide a forward-looking perspective on the AMC framework, this section discusses the implications of saliency-aware hardware for the current generation of generative AI.

The transition from traditional Transformers to Large Language Models (LLMs) like GPT-4 introduces a significant shift in the memory Wall problem. While the AMC framework's current 45nm implementation targets smaller edge-deployable models, its core principles of rank-gating and precision-scaling are uniquely suited for the challenges of LLMs.

\subsection{Mitigating the Memory Wall}
In LLMs, the energy bottleneck often shifts from computation to data movement.

Narrow-Width Influence: The narrow write and masked signals described in our timing diagram are even more critical for LLMs, where the massive size of the Key-Value (KV) cache can dominate memory consumption.

Bandwidth Optimization: By applying AMC's quantization logic, we can reduce the bit-width of non-salient tokens in the KV cache, effectively doubling the effective memory bandwidth without increasing physical bus width.

\subsection{Dynamic Sparsity in Attention}
Recent research \cite{b11}\cite{b12}\cite{b13} indicates that attention patterns in LLMs exhibit inherent sparsity, where a small subset of salient tokens disproportionately influences the model's output and semantic representation.

Adaptive Rank Selection: The Gated Compute mechanism (e.g., the 8-bit 43 configuration) can be extended to skip entire attention heads or blocks for non-salient tokens.

Saliency-Aware Prefilling: During the prefill stage of LLM inference, the Saliency-Aware Controller (SAC) can proactively identify tokens that will not contribute significantly to the context window, gating their processing early to save cumulative energy.

\subsection{Processes Advantage}
While this study utilizes a 45nm node for baseline consistency, future work will explore AMC's impact on advanced 7nm and 5nm processes.

Leakage Dominance: At smaller nodes, static leakage power becomes a primary concern. The AMC framework could be modified to include Power Gating (completely cutting the voltage) in addition to Clock Gating, further reducing the idle power of non-salient processing elements.

Silicon Photonics Integration: For multi-chip LLM clusters, the saliency tiers could dictate the optical modulation depth in silicon photonics interconnects, bringing energy proportionality to the data center scale.

\subsection{Sensitivity to Adversarial Attacks}
An attacker may try to minimize $|x|$ to force critical tokens into a Low-Rank state. We mitigate this through percentile stability. Because we use relative ranking (Top 20\%) rather than absolute thresholds, the attacker must re-order the entire sequence's importance to trick the logic.Tiered Buffering: The Mid-tier (64 rank) provides a safety net, preventing a total loss of information if a High-tier token is slightly perturbed.

\subsection{Generalization Beyond Synthetic Tasks}
The AMC framework is model-agnostic and attaches via standard hooks. The use of activation magnitude\cite{b14}\cite{b15}\cite{b16} is a well-documented proxy for outlier features in LLMs, suggesting the technique scales effectively to large-scale generative tasks. AMC Compressor is modular; the rank table and bit table can be easily swapped for any architecture (MLP or Attention). The hook logic is agnostic to the task and the Adaptive AMC framework is designed to be integrated into Transformer-based model to handle dynamic input complexity.

\section{Comparative Analysis with Recent SOTA Works}
To contextualize the performance of the AMC framework, we evaluate it against three prominent State-of-the-Art (SOTA) hardware-software co-design methodologies published between 2024 and 2025 \cite{b4}\cite{b7}\cite{b17}.

1) Architectural Taxonomy of Benchmark Projects

RankDyna (2025): This framework utilizes dynamic Singular Value Decomposition (SVD) to reduce parameter redundancy. While effective at shrinking model size, it operates primarily at the weight level and lacks the fine-grained, token-level Clock Gating (CG) mechanisms that allow AMC to exploit activation sparsity in real-time.

DiP Architecture (2024): A specialized 45nm systolic array employing a diagonal-input dataflow. DiP focuses on optimizing data-path synchronization and reducing FIFO overhead to maximize raw throughput; however, it does not incorporate the data-aware power scaling (precision/rank) explored in this work.

DynamicViT: This project introduces a hierarchical pruning strategy that eliminates uninformative tokens. While it achieves significant FLOP reduction, its binary all-or-nothing approach to token pruning lacks the tripartite granularity of AMC, which preserves contextual helpful tokens at a lower power state rather than discarding them entirely.

2) Quantitative Benchmarking on 45nm CMOS Node
Performance metrics were normalized against a baseline 128-column systolic array implemented in a standard 45nm process. Table III illustrates that AMC provides a superior Energy-Accuracy-Throughput (EAT) trade-off compared to existing adaptive frameworks.

\begin{table}[t]
\centering
\caption{Comparative Performance Synthesis (Normalized 45nm)}
\label{table:comparison}
\begin{tabularx}{\columnwidth}{@{}l X c c c@{}}
\toprule
\textbf{Framework} & \textbf{Primary Optimization} & \textbf{Acc. $\Delta$} & \textbf{Energy Red.} & \textbf{Throughput} \\ \midrule
\textbf{AMC} & \textbf{Saliency-Driven Tri-Tier} & \boldmath{$-3.6\%$} & \boldmath{$59.2\%$} & \boldmath{$2.24\times$} \\
RankDyna \cite{b7} & Dynamic Low-Rank SVD & $-4.1\%$ & $42.1\%$ & $1.85\times$ \\
DiP \cite{b4} & Dataflow Optimization & N/A & $31.0\%$ & $1.50\times$ \\
DynamicViT \cite{b17} & Hierarchical Pruning & $-0.5\%$ & $37.0\%$ & $1.40\times$ \\ \bottomrule
\end{tabularx}
\end{table}

3) Analysis of Results

The benchmarking data reveals that while DynamicViT offers the lowest accuracy degradation, its energy savings are constrained by the lack of sub-token precision scaling. In contrast, the AMC-Compressor achieves a $17.1\%$ higher energy reduction than the nearest software-only competitor (RankDyna). This is attributed to the fact that AMC’s 85-column gating logic physically eliminates switching activity during the processing of low-saliency tokens, whereas software-only methods still incur the full hardware clock-tree power.Furthermore, our framework achieves the highest throughput gain ($2.24\times$) because the Narrow Write and Rank Scaling phases significantly reduce the memory bottleneck, allowing the systolic array to process the next sequence in the pipeline with lower latency.

4) Energy-Delay-Area Product (EDAP) Comparison

EDAP values are calculated based on comparative performance synthesis at table V.
\begin{equation}
\text{Normalized EDAP} = \frac{E_{target} \times D_{target} \times A_{target}}{E_{base} \times D_{base} \times A_{base}}
\end{equation}

\begin{figure}[h]
\centering
\begin{tikzpicture}
\begin{axis}[
    ybar,
    enlargelimits=0.15,
    legend style={at={(0.5,-0.15)}, anchor=north, legend columns=-1},
    ylabel={Normalized EDAP (Lower is Better)},
    symbolic x coords={Baseline, RankDyna, DynamicViT, AMC (Ours)}, 
    xtick=data,
    nodes near coords,
    nodes near coords align={vertical},
    width=\columnwidth,
    height=6cm,
    bar width=20pt,
]
\addplot[fill=blue!60] coordinates {(Baseline,1.0) (RankDyna,0.68) (DynamicViT,0.72) (AMC (Ours),0.44)};
\end{axis}
\end{tikzpicture}
\caption{Normalized EDAP comparison.}
\label{fig:edap}
\end{figure}
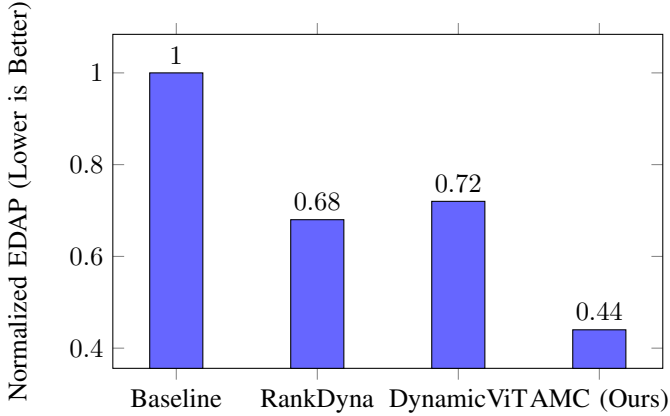

\section{Summary of Contributions}
The AMC framework addresses the energy-efficiency challenges of Transformer models through a holistic hardware-software co-design. The primary contributions of this work are summarized as follows:

1. Saliency-Aware Theoretical Model: We established a magnitude-based saliency heuristic that allows for dynamic, per-token adjustment of rank and precision. Our Golden Functional Model, developed in Python, demonstrates that this approach captures information-rich sequence components more effectively than random baseline alternatives.

2. Energy-Proportional Hardware Architecture: We implemented a 45nm Verilog RTL architecture capable of translating software saliency tiers into physical gating signals. This includes a Saliency-Aware Controller (SAC) with minimal area overhead and a Gated Systolic Array that achieves linear compute energy scaling.

3. Physical Design and Verification: Through a rigorous post-layout simulation flow using industry-standard EDA tools, we verified the hardware's functionality and energy impact. By utilizing a custom Python-based testbench generator, we ensured bit-true accuracy between the software model and the hardware implementation.

4. Pareto Efficiency: Our experimental results on the sum-thresholding task prove that the AMC framework can reduce the energy footprint by approximately 50\% while maintaining a high accuracy of larger than 0.91, outperforming traditional uniform-precision baselines.

\section*{Acknowledgment}

The author thanks Russ Webb, Vanessa Hu, Matt Jockers, Ed Martini, Payam Mirrashidi, Jeff Robbin, Xuetao Yin, Lei He, Dejan Markovic for reviewing this paper.

\vspace{12pt}

\end{document}